\documentclass[12pt]{iopart}

\newcommand{\unit}[1]{\ensuremath{\;\mathrm{#1}}}

\usepackage{graphicx}
\usepackage{hyperref} 
\hypersetup{bookmarks=true,colorlinks=true,linkcolor=blue,citecolor=green}

\begin{document}


\title{Entangled Quantum Key Distribution with a Biased Basis Choice}

\author{Chris Erven,$^1$ Xiongfeng Ma,$^1$ Raymond Laflamme,$^{1,2}$ and Gregor Weihs$^{1,3}$}
\address{$^1$ Institute for Quantum Computing and Department of Physics and Astronomy, University of Waterloo, 200 University Avenue West, Waterloo, ON, N2L 3G1, Canada}
\address{$^2$ Perimeter Institute, 31 Caroline Street North, Waterloo, ON, N2L 2Y5, Canada}
\address{$^3$ Institut f\"ur Experimentalphysik, Universit\"at Innsbruck, Technikerstrasse 25, 6020 Innsbruck, Austria}
\ead{cerven@iqc.ca}

\date{\today}

\begin{abstract}
We investigate a quantum key distribution (QKD) scheme which utilizes a biased basis choice in order to increase the efficiency of the scheme. The optimal bias between the two measurement bases, a more refined error analysis, and finite key size effects are all studied in order to assure the security of the final key generated with the system. We then implement the scheme in a local entangled QKD system that uses polarization entangled photon pairs to securely distribute the key. A 50/50 non-polarizing beamsplitter with different optical attenuators is used to simulate a variable beamsplitter in order to allow us to study the operation of the system for different biases. Over 6 hours of continuous operation with a total bias of 0.9837/0.0163 (Z/X), we were able to generate 0.4567 secure key bits per raw key bit as compared to 0.2550 secure key bits per raw key bit for the unbiased case. This represents an increase in the efficiency of the key generation rate by 79\%.
\end{abstract}



\section{Introduction}\label{sec.Introduction}

Quantum key distribution (QKD) allows two distant parties, Alice and Bob, to create a random secret key even when the quantum channel they share is accessible to an eavesdropper, Eve, so long as they also have an authenticated public classical channel. The security of QKD is built on the fundamental laws of physics in contrast to existing classical public key cryptography whose security is based on unproven computational assumptions.

There are mainly two types of QKD schemes: prepare-and-measure schemes, the best known of which is the original BB84 protocol proposed by Bennett and Brassard \cite{BB84} in 1984; and entanglement based schemes, the simplest of which is the BBM92 scheme developed by Bennett \emph{et al.} \cite{BBM92} in 1992. The BBM92 scheme essentially symmetrizes the BB84 protocol by distributing entangled pairs of qubits to Alice and Bob and having them both measure their half of each pair in one of two complementary bases. For a more complete overview of both QKD theory and experiments, please refer to the recent review articles by Gisin \emph{et al.} \cite{GRTZ02}, {Du\u sek} \emph{et al.} \cite{DLH06}, and Scarani \emph{et al.} \cite{SBCDLP08}.

A key feature of the BB84 protocol and many others is that the bases used are chosen randomly, independently, and uniformly. Most security proofs, including the seminal work by Shor and Preskill \cite{SP00}, rely heavily on the symmetry which uniformity of basis choice provides. For example, it allows the sifted data from both bases to be grouped together and a simple error correction algorithm to be performed on the grouped data producing a single error rate. However, uniformly chosen bases have the consequence that on average half of the raw data is rejected leading to an efficiency limited to at most 50\%. This is the reason for the curious factor of $\frac{1}{2}$ that appears in the key rates of many security proofs \cite{Lut00,MFL07b}.

This symmetry requirement was removed by Lo \emph{et al.} \cite{LCA02} in 2004 when they proposed a simple modification to the BB84 scheme that could in principle allow one to asymptotically approach an efficiency of 100\%. Their scheme relies on two changes to the BB84 protocol: non-uniformity in the choice of bases, and a refined data analysis. The first change of non-uniformity allowed Alice and Bob to achieve much higher efficiencies with their raw data. In fact, Lo \emph{et al.} showed that this efficiency could be made arbitrarily close to 100\% in the long key limit. The second change of a refined data analysis allowed Alice and Bob to maintain the security of their system since a simple error analysis was no longer sufficient. Ac\'{\i}n \emph{et al.} \cite{AMP06} have also studied this protocol in 2006 using a CHSH test for security under the conditions of no-signalling eavesdroppers.

In this article, we detail the experimental implementation of the biased basis choice protocol with a simulated variable non-polarizing beamsplitter and a local entangled QKD system. We begin by first reviewing the theory for the biased protocol. We study the optimal bias ratio and the important parameters necessary to maintain security. We then follow with a description of the experimental setup for the implementation of the biased scheme. Lastly, we report on the results of the experiment and compare the efficiency of the biased protocol with those of an unbiased protocol.

\section{Theory}\label{sec.Theory}

Lo \emph{et al.} \cite{LCA02} proposed their protocol as a modification to the original BB84 protocol which is a prepare-and-measure scheme. Here we make the simple extension to the entanglement based BBM92 scheme developed by Bennett \emph{et al.} \cite{BBM92} in 1992. In the original scheme, a source of polarization entangled photon pairs in the singlet Bell state is placed in between Alice and Bob, and one photon from each pair is sent to Alice and Bob. Alice and Bob then randomly, independently, and uniformly choose to measure in either the rectilinear (H/V) basis or the diagonal ($+45^{\circ}$/$-45^{\circ}$) basis. Now we extend this scheme to remove the uniformity in the basis choices just as Lo \emph{et al.} did for the BB84 protocol. We are allowed to do so without violating the security of the scheme because the proof by Lo \emph{et al.} already relies on expanding the BB84 scheme into an imagined entanglement based scheme. Thus, their proof of security holds for performing the actual biased entanglement based scheme as well as performing the biased BB84 scheme. Additionally, we use the recently developed squashing model \cite{BML08b, TT08b, KAYI08} which allows us to assume that we are dealing with qubits so that the proof by Lo \emph{et al.} holds (double clicks are assumed to be rare and ignored).

The biased basis scheme has two main changes: first, Alice and Bob now choose their measurement bases randomly and independently but now non-uniformly with substantially different probabilities. This allows for a much higher probability of Alice and Bob using the same basis and thus allows them to achieve much higher efficiencies with their raw data. With uniformity removed, the second change necessary is a refined error analysis since an eavesdropper could now easily break a system which performed a simple error analysis on the lumped data by eavesdropping primarily along the predominant basis. To ensure security, it is crucial for Alice and Bob to divide their data into two subsets according to the bases used and compute an error rate for each subset separately. It is only with this second addition that one can ensure the security of this biased scheme.

First we define the necessary quantities that we will use in our security analysis. Define $e_{bx}$ $(\delta_{px})$ and $e_{bz}$ $(\delta_{pz})$ to be the bit (phase) error rates in the $X$ (diagonal) and $Z$ (rectilinear) bases respectively, where an $e$ is used to denote a measurable quantity and a $\delta$ is used to denote a quantity that has to be inferred. Note that the bit error rates, $e_{bx}$ and $e_{bz}$, are known exactly by Alice and Bob once they perform error correction since they can count the number of errors found during error correction. However, the phase error rates, $\delta_{px}$ and $\delta_{pz}$, need to be estimated from the bit error rates since they are not directly accessible from Alice and Bob's measurement data.

In order to estimate the phase error rates, we define the quantities $p_{bx}$ $(p_{px})$ and $p_{bz}$ $(p_{pz})$ to be the bit (phase) error probabilities in the $X$ and $Z$ bases respectively. Since a basis independent source is assumed, we know that
\begin{equation} \label{eq.Probs}
    p_{pz} = p_{bx} \qquad p_{px} = p_{bz}.
\end{equation}
Now in the long key limit $e_{bx}$ converges to $p_{bx}$ while $\delta_{pz}$ converges to $p_{pz}$. Using Eq. (\ref{eq.Probs}) now allows us to say that
\begin{equation}\label{eq.PhaseEqualsBit}
    \delta_{px} = e_{bz} \qquad \delta_{pz} = e_{bx}
\end{equation}
in the long key limit. Obviously for finite key lengths this equality will not hold exactly and we will have to consider statistical fluctuations. We will address the finite key size effects in a moment, but for now assume that Eq. (\ref{eq.PhaseEqualsBit}) holds.

The key point of the security analysis rests in the privacy amplification part \cite{BBCM95}. Privacy amplification is usually performed via the 2-universal hash functions discovered by Wegman and Carter \cite{CW79} and that is what is done in the experimental implementation detailed in this paper. Alice and Bob need to take care of the bits exposed during error correction which they can directly count during the error correction process since an eavesdropper can learn this information listening in to the classical channel. They also need to estimate the phase error rates for the two bases in order to estimate the amount of an eavesdropper's information on the quantum signals that were distributed. Privacy amplification then needs to be applied to reduce the information of the eavesdropper from these two sources to an arbitrarily small amount. This leads to the following key generation rate according to \cite{MFL07b}, expressed in terms of secure bits per raw bit, using Eq. (\ref{eq.PhaseEqualsBit}) above
\begin{eqnarray}\label{eq.KeyRate}
    R & \geq & (1-q)^2[1 - f(e_{bx})h_{2}(e_{bx}) - h_{2}(\delta_{px})] \nonumber \\
     & & + q^2[1 - f(e_{bz})h_{2}(e_{bz}) - h_{2}(\delta_{pz})] \nonumber \\
    & \geq & (1-q)^2[1 - f(e_{bx})h_{2}(e_{bx}) - h_{2}(e_{bz})] \nonumber \\
     &  & + q^2[1 - f(e_{bz})h_{2}(e_{bz}) - h_{2}(e_{bx})]
\end{eqnarray}
where $q$ is defined to be Alice's or Bob's bias probability of measuring in the $Z$ basis and $(1-q)$ is their probability of measuring in the $X$ basis, $f(x)$ is the error correction inefficiency as a function of the error rate, normally $f(x) \geq 1$ with $f(x) = 1$ at the Shannon limit, and $h_{2}(x) = -x \log x - (1-x) \log (1-x)$ is the binary entropy function. For this initial analysis, the key rate formula assumes that Alice and Bob each pick the same identical bias.

We benefit from separating the key rate into contributions from the $X$ and $Z$ bases since the key rate can still be positive for an error rate higher than 11\% in one basis so long as the other error rate is low enough. This can be seen in Fig. 2 of Ref. \cite{MFDCTL06} since we use local operations and one way classical communication (1-LOCC) in our post-processing. Note that the cascade error correction algorithm, which we use, is considered 1-LOCC post-processing since the two way classical communication is not used to perform advantage distillation. Also note that in this treatment of cascade we assume that Eve learns both the positions of the errors and the revealed parity bits.

The last thing to take care of are the finite key size effects since these will be very important in determining how to balance the optimal biasing ratio. For this analysis we assume the main finite key size effect is due to the parameter estimation; namely, the statistical fluctuations in the phase error rates compared to what is estimated from the bit error rates. The other possible finite key size effects are authentication, the leakage of information during error correction (typically referred to as $\mathrm{leak_{EC}}$), the probability of failure for error correction and error verification (typically referred to as $\varepsilon_{EC}$), and the probability of failure of privacy amplification (typically referred to as $\varepsilon_{PA}$) \cite{SR08b,CS08}.

For our experiment the probability of failure for our parameter estimation is on the order of $10^{-6}$ since, as will be discussed below, we choose a safety margin on our phase error estimates so that the probability that the actual phase error rates are outside of this range is less than $10^{-6}$. Also, we do not implement authentication on the classical channel, so we ignore its effect on our key rate. Note though that the resources required for authentication scale logarithmically in the length of the secret key generated by a QKD session \cite{ABBDDGGGLLLPPPPRRRRSSWZ07}. The information being leaked during error correction does not contribute any finite key size effects since we directly count the number of bits revealed and take care of them in the privacy amplification step \cite{Lut99}. Thus, there are no fluctuations in this contribution since we know it exactly. The error correction and verification algorithm we implement is due to Sugimoto and Yamazaki \cite{SY00} and allows us to bound the failure probability and thus set its value. For this experiment we implement the error correction algorithm so that its probability of failure is $< 10^{-10}$. Lastly, the probability of failure for the privacy amplification step is assumed to be negligible since it depends on the privacy amplification algorithm used and we assume that its possible to make one with a sufficiently small failure probability. For a strict finite key analysis, one should follow Refs. \cite{HHHT07,SR08,SR08b,CS08}.

We need to consider the statistical fluctuations in order to discuss the optimal bias between the two bases. As was stated above, both error rates $e_{bx}$ and $\delta_{pz}$ converge to $p_{bx} = p_{pz}$ in the long key limit. We use standard random sampling theory to determine the following formula given in \cite{MFL07b} for the estimates of our phase error rates
\begin{eqnarray} \label{eq.RandomSampling}
    P_{\epsilon_{z}} & \equiv & \mathrm{Prob} \{\delta_{pz} > e_{bx}+\epsilon_{z}\} \nonumber \\
    & \leq & \exp[-\frac{\epsilon_{z}^2 n_{xx}}{4 e_{bx} (1-e_{bx})}],
\end{eqnarray}
where $n_{xx}$ is the number of bits generated from Alice and Bob both measuring in the $X$ basis, and $\epsilon_{z}$ is a small deviation from the measured bit error rate. Eq. (\ref{eq.RandomSampling}) allows us to put a bound on the probability that the phase error rate deviates from our measured bit error rate by more than $\epsilon_{z}$. For example, if Alice and Bob measure 10,000 qubits in the $X$ basis ($n_{xx} = 10,000$), find an error rate of 5\% ($e_{bx} = 0.05$), and set their desired deviation to 1\% ($\epsilon_{z} = 0.01$), then Eq. (\ref{eq.RandomSampling}) would allow them to say that the probability that their phase error rate ($\delta_{pz}$) deviates by more than $\epsilon_{z}$ is less than 0.52\%. Or said another way, if we use $\delta_{pz} = 0.06$ in our key rate formula then we are confident that our key was generated securely with a probability of 99.48\%. So our key rate formula has now become
\begin{eqnarray}\label{eq.KeyRateWithEpsilon}
    R & \geq & (1-q)^2[1 - f(e_{bx})h_{2}(e_{bx}) - h_{2}(e_{bz} + \epsilon_{x})] \nonumber \\
     & & + q^2[1 - f(e_{bz})h_{2}(e_{bz}) - h_{2}(e_{bx} + \epsilon_{z})],
\end{eqnarray}
which is the same as before except for the addition of $\epsilon_{x}$ and $\epsilon_{z}$ which are now necessary to deal with the finite key statistics. We should note that the random sampling formula used in Eq. (\ref{eq.RandomSampling}) is a good approximation in the long key limit.

\begin{figure}[hbt]
    \centering
    \includegraphics[width=15cm]{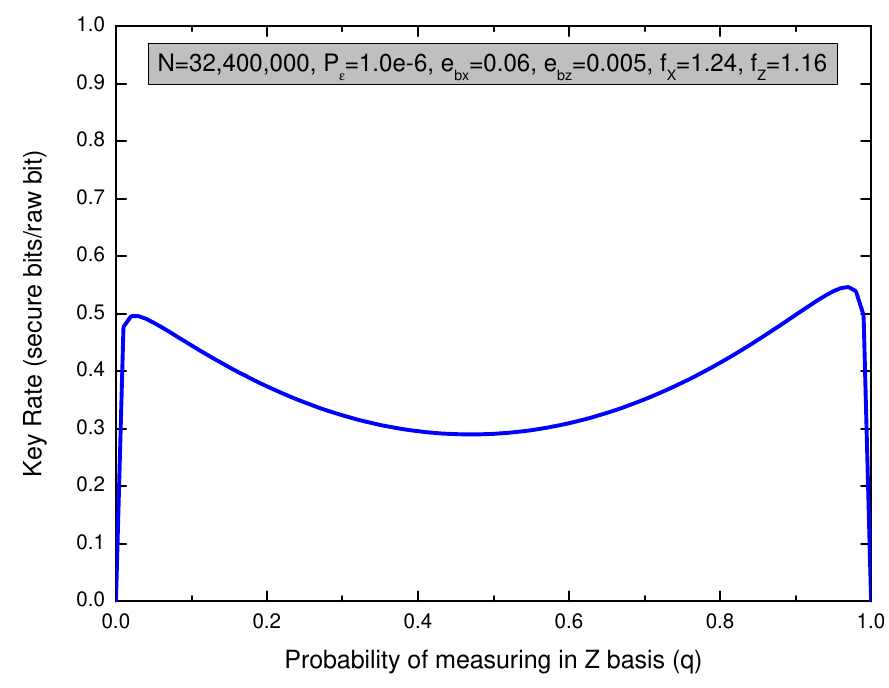}
    \caption{Plot of the key generation rate ($R$) in terms of the bias ratio ($q$).} \label{fig:Rr}
\end{figure}

Now we can try to find the optimal bias between the two bases. Given estimates for $e_{bx}$, $e_{bz}$, and $N$ (the total coincidence count), and picking $P_{\epsilon} = P_{\epsilon_{x}} + P_{\epsilon_{z}}$, one can optimize the bias $q$ and the deviations $\epsilon_{x}$ and $\epsilon_{z}$ according to Eqs. (\ref{eq.KeyRateWithEpsilon}) and (\ref{eq.RandomSampling}). For example, with estimates of the parameters we expect to see in our system, we can graph the key rate (normalized in terms of secure key bits per raw key bit) versus the bias ratio $q$, shown in Figure \ref{fig:Rr}. The inset shows the estimates for the parameters needed in order to find the optimum bias: $N$ is the total number of entangled pairs sent to Alice and Bob over the course of the experiment, $P_{\epsilon}$ is the confidence probability desired for our phase error statistics, $e_{bx}$ and $e_{bz}$ are the observed bit error rates in the $X$ and $Z$ bases over the course of the experiment, and $f$ is the observed error correction efficiency. From Fig. \ref{fig:Rr} we can see that the key rate is maximized for a bias of $q = 0.97$. Once we have found the optimum bias we can then use it, along with Eq. \ref{eq.RandomSampling}, to figure out the optimum deviations $\epsilon_{x}$ and $\epsilon_{z}$ that will still allow us to achieve our desired confidence probability $P_{\epsilon}$.

Examining Fig. \ref{fig:Rr} we see that a maximum also occurs around $q = 0.03$ though it is slightly lower than the one at $q = 0.97$. It is obvious that the efficiency curve should be symmetric about the middle point $q = 0.5$ since biasing the protocol towards the $X$ basis should be just as good as biasing it towards the $Z$ basis. However, the curve is not entirely symmetric because the error rates and error correction efficiencies are not identical in the two bases. The error correction efficiency plays the largest part in the overall rate since it costs a fraction of $f(e)h(e)$ in the final key generation rate. Thus, since the error correction efficiency is lower in the $X$ basis than the $Z$ basis ($f_{X} > f_{Z}$), the optimum rate is also lower when the bias is skewed towards the $X$ basis.

It is important to understand how the biased protocol utilizes the optimum bias in order to make the most efficient use of the raw key data. The optimum bias, $q$, along with the optimum deviations, $\epsilon_{x}$ and $\epsilon_{z}$, are chosen such that only the minimum number of measurements are made in the weak basis in order to achieve the desired confidence probability $P_{\epsilon}$. This has the consequence that privacy amplification of the measurement results from the strong basis has to be deferred until the end of the entangled photon distribution phase. It is only with the last distributed photon that Alice and Bob gain enough statistics in the weak basis to allow them to privacy amplify all the error corrected key generated in the strong basis over the course of the experiment. They will obtain small amounts of privacy amplified key from the weak basis over the course of the experiment, but the majority of the key that comes from the strong basis will not be available until after the distribution phase is completed.

\section{Experimental Implementation}\label{sec.ExperimentalImplementation}

The purpose of this experiment was the experimental investigation of the biased basis QKD protocol and all of the practicalities associated with implementing the protocol; such as, choosing the proper bias, doing a more refined error analysis, and worrying about the finite key size effects. In order to focus on these issues, we did not involve the complications of a free-space link and instead chose for Alice and Bob to locally detect each of their halves of the photon pairs while sitting next to the source connected to it with a short optical fibre. Additionally, since we wanted to investigate many different biases we decided to simulate a variable non-polarizing beamsplitter since they do not exist to the best of the author's knowledge. Indeed, even fixed biased non-polarizing beamsplitters are extremely expensive and difficult to manufacture; therefore, it is worthwhile to study the performance of the system for many different biases with a simulated variable biased beamsplitter first. The difficulties associated with biased beamsplitters might suggest one possible reason for employing an active basis selection scheme over a passive one since biasing might be done more easily. However, the active schemes would have their own complications for generating truly random, biased bit strings at a high enough rate for their polarization modulators. We are also currently investigating development of a variable non-polarizing beamsplitter to allow the flexible adjustment of the bias in our system without throwing away counts.

The experimental setup consists of a compact spontaneous parametric down-conversion (SPDC) source, two compact passive polarization analysis modules, avalanche photodiode (APD) detectors, time-stampers, GPS time receivers, two laptop computers, and custom written software. The SPDC source is comprised of a 1\unit{mm} thick $\beta$-BBO crystal pumped with a 50\unit{mW} laser which produces entangled photon pairs at a degenerate wavelength of 815\unit{nm}. A $\beta$-BBO crystal 0.5\unit{mm} thick in each arm compensates for walk-off effects, which can ruin the entanglement. Typically, a total single photon count rate of 100,000\unit{s^{-1}} in each arm of the source and a coincident detection rate of 11,000\unit{s^{-1}} is measured locally. More details on the setup can be found in our earlier papers \cite{WE07,ECLW08b,ECLW08}.

\begin{figure}[hbt]
    \centering
    \includegraphics[width=7cm]{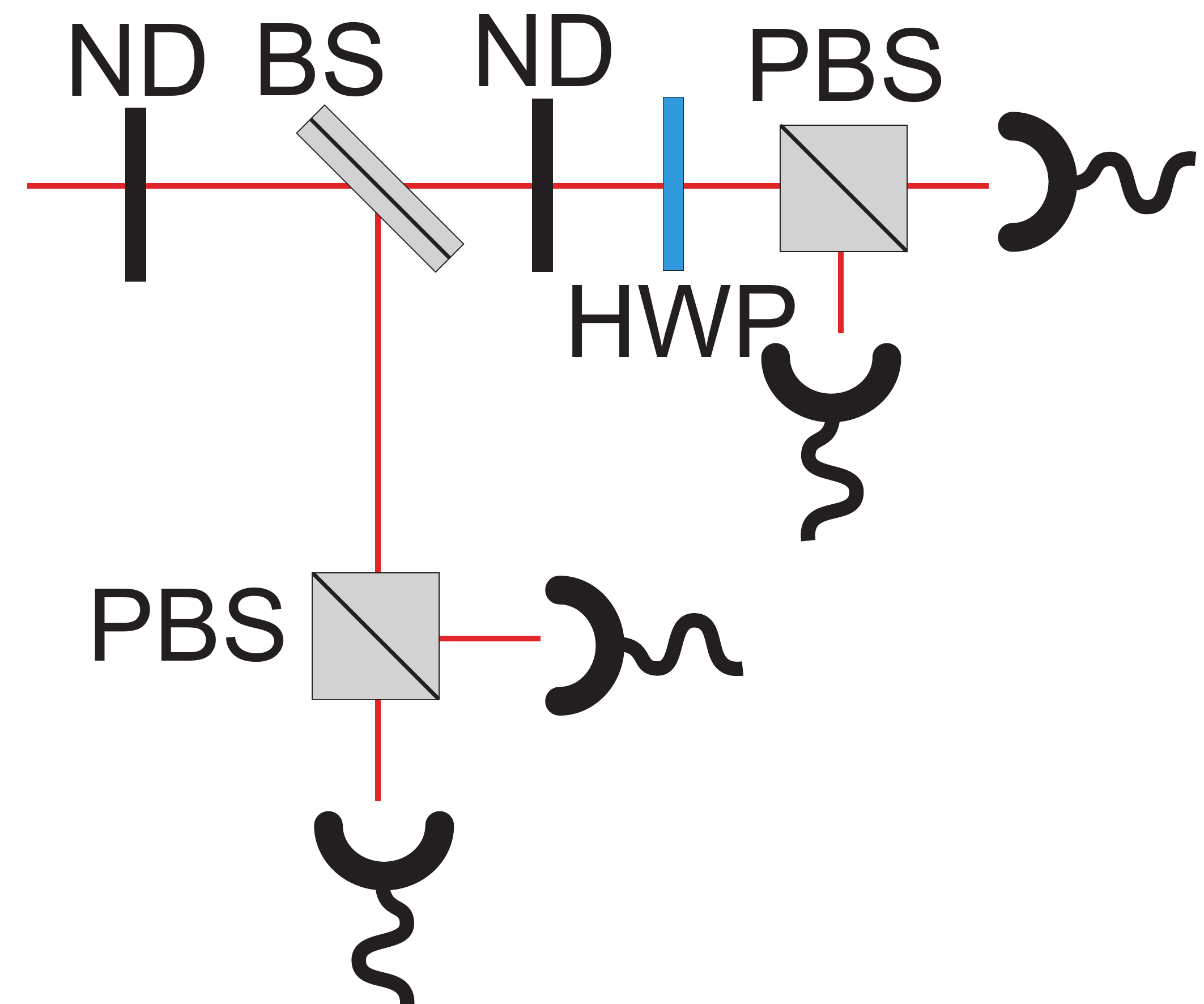}
    \caption{Schematic of the polarization analysis module with a neutral density (ND) filter placed in the transmission arm of the 50/50 beamsplitter (BS) in order to achieve the desired bias in the measurement results. (Polarizing beamsplitters (PBS))} \label{fig:PolAnaMod}
\end{figure}

In order to implement the biased protocol we simulated a biased beamsplitter by placing the appropriate attenuators in the transmission arm of the original 50/50 beamsplitter (BS) used to perform the basis choice as shown in Fig. \ref{fig:PolAnaMod}.
The transmitted arm analyzes the photons in the diagonal basis. Also, in order to directly compare the efficiencies of experiments with different biases we used the appropriate attenuators ahead of the beamsplitter to make the rates of each experiment with a different bias equal. The attenuators are placed in the transmission arm ($X$, diagonal basis) so that the $Z$ (rectilinear) basis is the one which Alice and Bob predominantly measure in. As was mentioned earlier, we make this choice because error correction costs a factor of $f(e)h(e)$ (with $f(e) > 1$) while privacy amplification costs a factor of $h(e)$ in our key generation rate. Thus, since the $Z$ (rectilinear) basis has a much lower intrinsic error rate, we would like to make it the predominant basis since less error correction will be needed than if we chose the $X$ (diagonal) basis to be the predominant one.

Our custom written software needed to be modified in order to analyze the error rates in both bases separately and defer the privacy amplification until the end of the entangled pair distribution phase.  Error correction is first performed on the $X$ basis measurements followed by the $Z$ basis measurements, revealing the bit error rates $e_{bx}$ and $e_{bz}$. The number of bits revealed during the error correction of the $X$ and $Z$ measurements are recorded. After the distribution phase, we use the actual experimental results for $N$, $n_{xx}$, $n_{zz}$, $q$, $e_{bx}$, $e_{bz}$, $f(e_{bx})$, and $f(e_{bz})$, along with the desired $P_{\epsilon_{x}}$ and $P_{\epsilon_{z}}$ to calculate the optimum $\epsilon_{x}$ and $\epsilon_{z}$ according to Eq. \ref{eq.RandomSampling}. This allows us to distill the maximum amount of key from our raw data. The appropriate privacy amplification factor, according to Eq. \ref{eq.KeyRateWithEpsilon}, is now calculated for each measurement set. Privacy amplification using a 2-universal hash function \cite{BBR88,BBCM95,CW79} then takes care of the bits revealed during error correction, the estimated phase error rates, and the additional safety margin needed for the finite key statistics to produce the final secure key.

For this experiment we improved our error correction algorithm since knowing the bit error rates $e_{bx}$ and $e_{bz}$ as precisely as possible is extremely important. To do this we improved the cascade error correction algorithm \cite{BBSBS92,BS94}, which was initially used in a modified form in our system, to the optimized algorithm outlined by Sugimoto and Yamazaki \cite{SY00}. In previous experiments with our original modified cascade algorithm we achieved a residual bit error rate of $1.92 \times 10^{-3}$ \cite{ECLW08}, which is clearly insufficient for realistic key sizes especially once finite key size effects have to be taken into account. In order to be secure, it has been shown that privacy amplification needs to work with key lengths on the order of $\sim 10^{7}$ bits \cite{CS08}. Consequently, the probability of residual errors needs to be a least two orders of magnitude less than this in order for the privacy amplification to succeed with high probability. With the improved optimized cascade algorithm we are now able to set a parameter, $s$, which then determines the residual bit error rate according to $P_{residual} < 2^{-s}$. For this experiment we chose $s = 40$ which should give us a maximum residual bit error rate of $9.09 \times 10^{-13}$.

\section{Results}\label{sec.Results}

On the day of the experiment, we measured visibilities of 99.6\% and 92.4\% are directly measured in the rectilinear and diagonal bases respectively. This corresponds to baseline error rates in the two bases of $e_{bx} = 0.038$ and $e_{bz} = 0.002$ due to the source. The limited visibility in the diagonal basis (high $e_{bx}$) is likely due to the broad spectral filtering (10\unit{nm}) in the polarization detector box which allows some unentangled photon pairs through the filters which still have a strong correlation in the rectilinear basis, but have almost no correlation in the diagonal basis. The limited visibility is also likely due to uncompensated transverse walk-off in the $\beta$-BBO crystal which is aggravated by the narrow pump beam spot.

\begin{figure}[hbt]
    \centering
    \includegraphics[width=16cm]{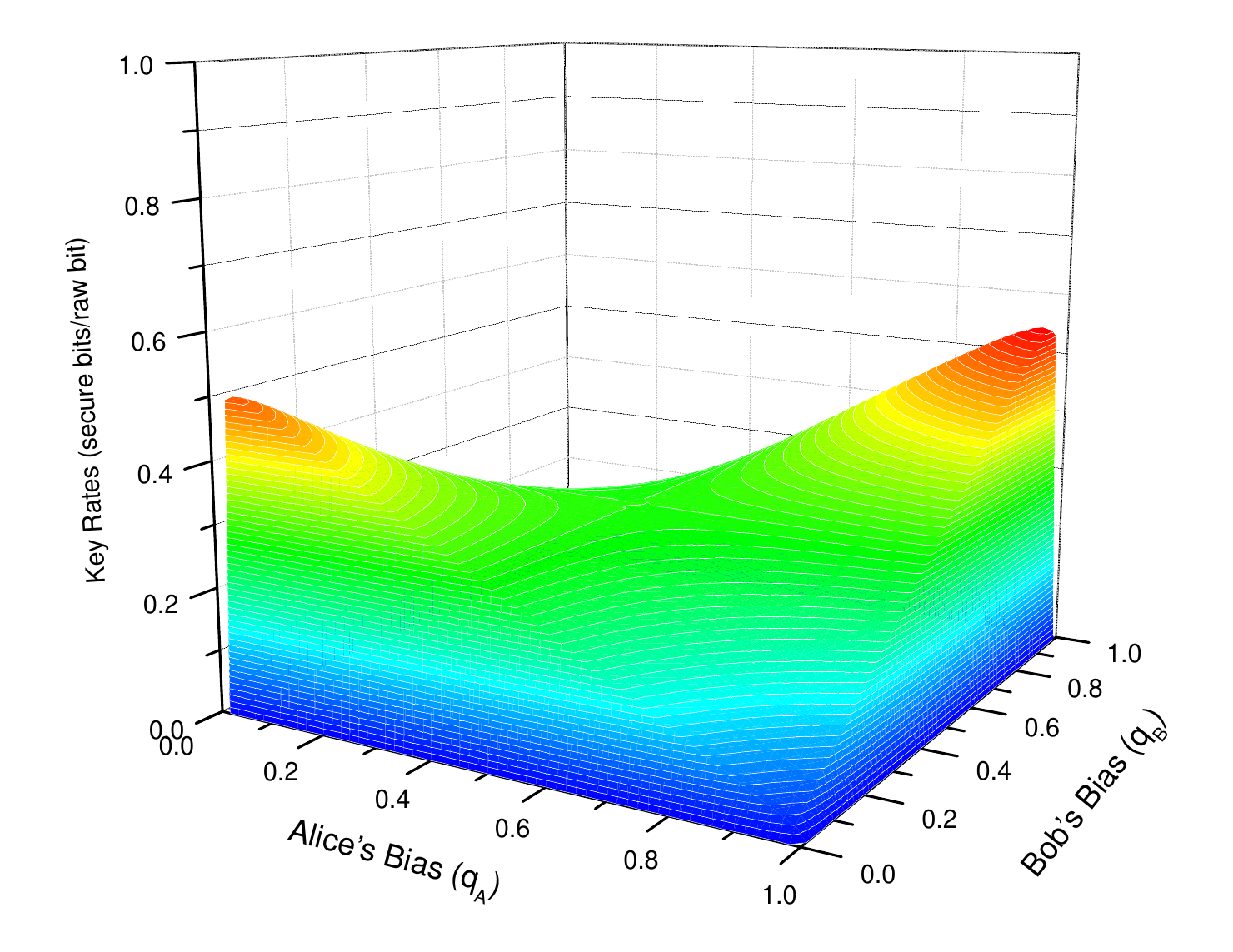}
    \caption{Plot of the key generation rate (R) in terms of Alice's bias ratio ($q_{A}$ and Bob's bias ratio ($q_{B}$). Alice's bias is plotted along the left axis while Bob's is plotted along the right.} \label{fig.BiasVsKeyRate3D}
\end{figure}

We performed four different experiments with the varying biases shown in Table \ref{tab.Biases}, each approximately six hours in duration in order to compare their efficiencies. While the appropriate attenuators were put into the transmission arms of both Alice's and Bob's detectors to simulate the desired bias, differences in coupling efficiencies within the polarization analysis modules produced slightly asymmetric biases between Alice and Bob for the $Z$ basis choice. In order to figure out the optimum $\epsilon$'s needed and the proper privacy amplification factor, the simple uniform bias analysis above had to be expanded into a more complex analysis that allowed Alice and Bob to have non-identical biases. Fig. \ref{fig.BiasVsKeyRate3D} is the generalization of Fig. \ref{fig:Rr} and plots the key rate (R) versus Alice's bias (shown along the left axis) versus Bob's bias (shown along the right axis). Using this analysis we proceeded to find the optimum $\epsilon$'s, calculate the privacy amplification factors, and complete the key generation process.

\begin{table}[htbp]
  \centering
  \begin{tabular}{|c|c|c|c|}
      \hline
      Exp & \multicolumn{3}{|c|}{Bias (Z/X)} \\
      \cline{2-4}
      \# & Alice & Bob & Total \\
      \hline
      1 & 0.4570/0.5430 & 0.4752/0.5248 & 0.4343/0.5657 \\
      \hline
      2 & 0.5660/0.4340 & 0.6074/0.3926 & 0.6639/0.3361 \\
      \hline
      3 & 0.7398/0.2602 & 0.7606/0.2394 & 0.8938/0.1062 \\
      \hline
      4 & 0.8804/0.1196 & 0.9062/0.0938 & 0.9837/0.0163 \\
      \hline
  \end{tabular}
  \caption{The observed biases in each of the experiments. \label{tab.Biases}}
\end{table}

\begin{figure}[hbt]
    \centering
    \includegraphics[width=18cm]{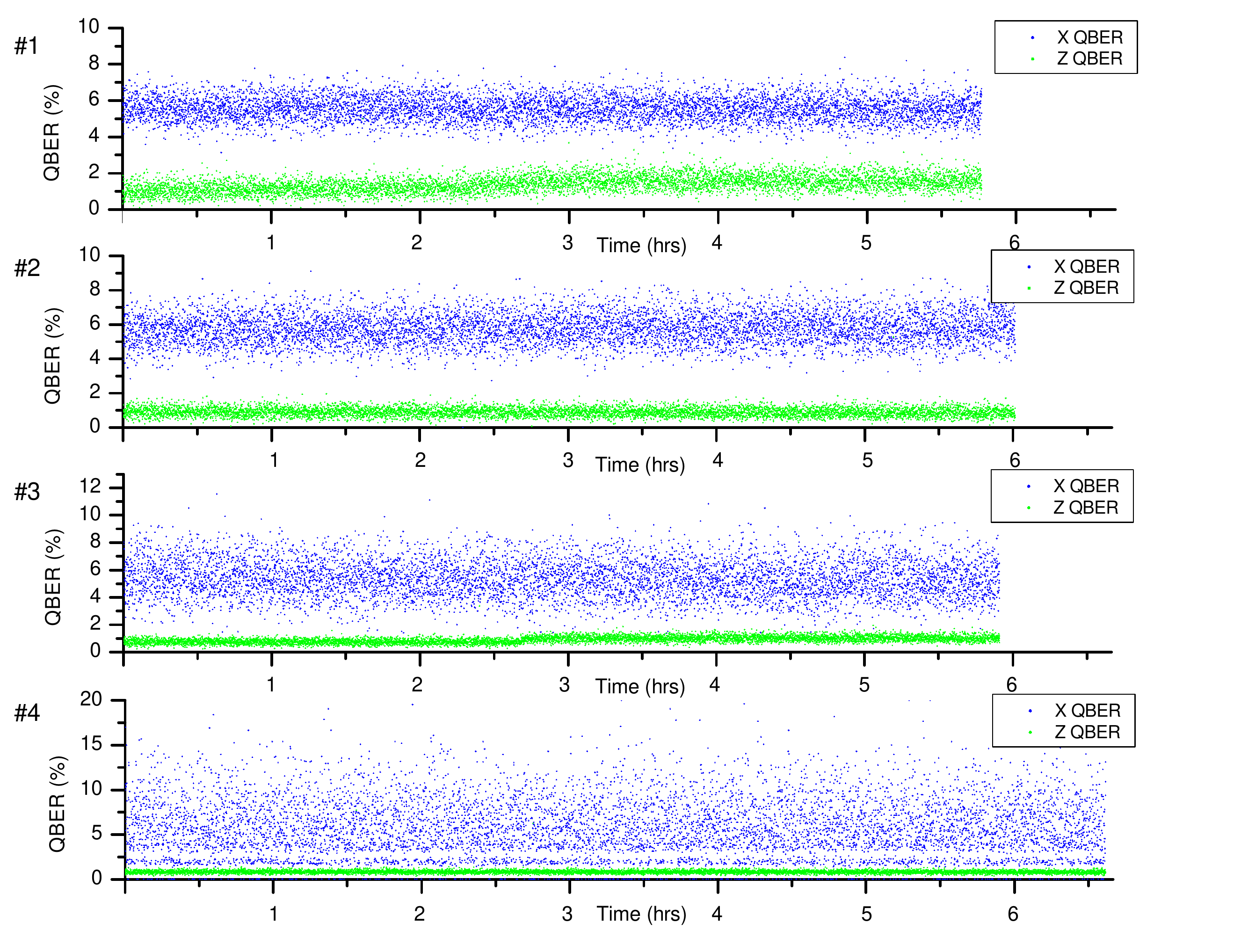}
    \caption{Plot of the QBER's in the X (blue) and Z (green) bases over the course of the experiments.} \label{fig.QBER}
\end{figure}

Fig. \ref{fig.QBER} shows the QBER's in the $X$ and $Z$ bases measured over the course of each experiment. The average QBER's in the $X$ and $Z$ bases for each experiment are tabulated in Table \ref{tab.Results}. The increase in the QBER's from the baseline 3.8\% and 0.2\% to those observed is attributed to the typical leakage of the polarizing beamsplitters in the polarization analysis modules, the uncompensated birefringence in the singlemode fiber used to transport the photons between the source and the polarization analysis modules, and to accidental coincidences.

\begin{table}[htbp]
  \centering
  \begin{tabular}{|c|c|c|c|c|c|c|c|}
      \hline
      Exp & \multicolumn{2}{|c|}{QBER} & \multicolumn{3}{|c|}{Key Length} & Secure Bits & Inefficiency\\
      \cline{2-6}
      \# & X & Z & Raw & Sifted & Final & Per Raw Bit & \\
      \hline
      1 & 1.39\% & 5.55\% & 28,655,075 & 14,779,423 & 7,365,984 & 0.2550 & - \\
      \hline
      2 & 0.90\% & 5.76\% & 29,705,827 & 15,713,427 & 8,392,528 & 0.2825 & 1.11 \\
      \hline
      3 & 0.89\% & 5.36\% & 29,319,830 & 18,627,251 & 10,568,944 & 0.3605 & 1.41 \\
      \hline
      4 & 0.82\% & 5.80\% & 32,162,313 & 26,154,132 & 14,687,016 & 0.4567 & 1.79 \\
      \hline
  \end{tabular}
  \caption{The QBERs and key rates for each of the experiments. \label{tab.Results}}
\end{table}

\begin{figure}[hbt]
    \centering
    \includegraphics[width=18cm]{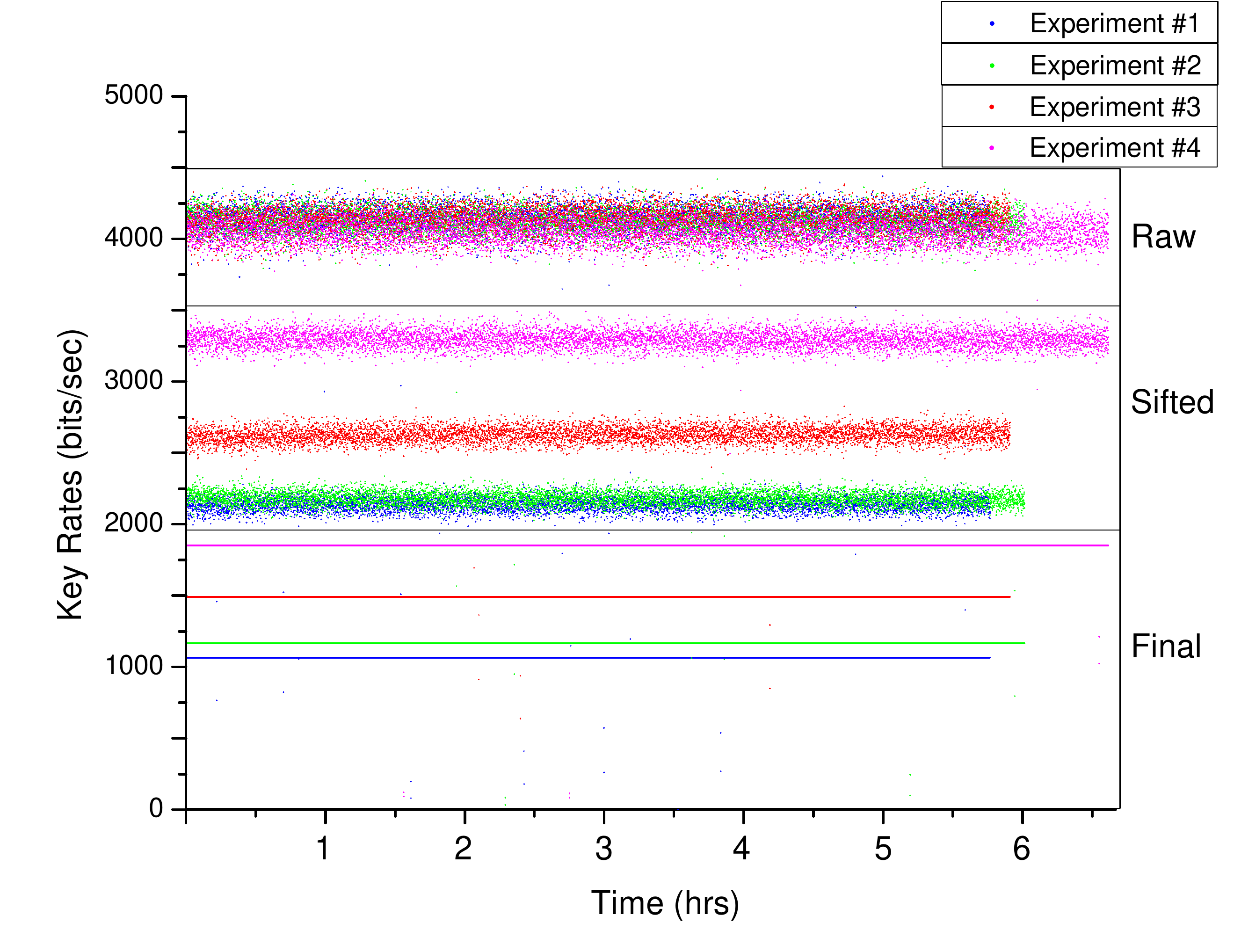}
    \caption{Plot of the raw, sifted, and average final key rates over the course of the experiment each experiment. The rates are grouped by colour with Experiment \#1 in blue, Experiment \#2 in green, Experiment \#3 in red, and Experiment \#4 in magenta. The upper box holds the raw key rates, the middle box holds the sifted key rates, and the lower box holds the average final key rates} \label{fig.KeyRates}
\end{figure}

Fig. \ref{fig.KeyRates} shows the raw key rate, sifted key rate, and average final key rate over the course of each experiment. The statistics for each experiment are grouped by the same colour, the upper box holds the raw key rates, the middle box holds the sifted key rates, and the lower box holds the average final key rates for each experiment. Note that we can only show an average final key rate since privacy amplification has to be deferred until the end of the entangled photon distribution phase and error correction phase, and then is performed on the ``entire'' sifted key at once \cite{Note1}. The results of each experiment are summarized in Tab. \ref{tab.Results} along with their efficiencies compared to the first experiment which we take as the ``unbiased'' experiment. As is clearly shown in Table \ref{tab.Results} the efficiency of each experiment increases with the bias reaching a value of 0.4567 secure key bits per raw bit for the final experiment which had a bias of 0.9837/0.0163. Thus, by implementing the biased QKD protocol we were able to increase the secure key generation rate by 79\% over the unbiased case. Clearly, the use of a biased protocol for the generation of secure key bits results in a more efficient use of the distributed entangled photon pairs and allows Alice and Bob to distill more secret key from the same number of distributed pairs than what an unbiased protocol would allow. Additionally, as the number of entangled photon pairs is increased, the number of secure final key bits per raw key bit will approach 1.0 as was pointed out earlier.

\begin{table}[htbp]
  \centering
  \begin{tabular}{|c|c|c|}
      \hline
       \multicolumn{3}{|c|}{Average block sizes (X)} \\
      \hline
       Pass 1 & Pass 2 & Pass 3 \\
      \hline
       16 & 33 & 65 \\
      \hline
       \multicolumn{3}{|c|}{Average block sizes (Z)} \\
      \hline
       Pass 1 & Pass 2 & Pass 3 \\
      \hline
       72 & 144 & 289 \\
      \hline
  \end{tabular}
  \caption{Cascade stats - average block sizes. \label{tab.BlockSizes}}
\end{table}

Since the error correction algorithms were greatly improved during this experiment, we include actual data for their operation during experiment \#1. We use the error correction algorithm developed by Sugimoto and Yamazaki \cite{SY00} as an optimization of the cascade error correction algorithm developed by Brassard \emph{et al.} \cite{BS94} which was first mentioned in an earlier form in \cite{BBSBS92}. Table~\ref{tab.BlockSizes} shows the average block sizes used during the error correction of the sifted $X$ and $Z$ keys. Table~\ref{tab.ErrorSequence} shows the number of errors corrected during each pass and sequence of cascade. A pass is defined as a new random shuffling of the bits to form blocks of the sizes above which are then error corrected with BINARY \cite{BS94}. When an error is found, cascade then goes back through all previous sequences of shufflings of bits to correct errors that were missed beforehand. Table~\ref{tab.CascadeStats} shows the average numbers of errors corrected during the use of the BINARY and BICONF\cite{BS94} primitives and the corresponding number of bits revealed. Additionally, it shows the average key block lengths and QBERs found during cascade. Knowing these allows the calculation of the error correction efficiencies for our algorithm by first calculating the number of bits which an error correction algorithm operating at the Shannon limit would have revealed via $h_{2}(QBER) \times \mathrm{AvgKeyLength}$. These efficiencies are also shown in Table \ref{tab.CascadeStats} relative to an error correction algorithm operating at the Shannon limit.

\begin{table}[htbp]
  \centering
  \begin{tabular}{|c|c|c|c|c|}
      \hline
       \multicolumn{5}{|c|}{Average number of errors corrected (X)} \\
      \hline
       & Totals & Sequence 1 & Sequence 2 & Sequence 3 \\
      \hline
       Pass 1 & 31.4 & 31.4 & - & - \\
      \hline
       Pass 2 & 27.2 & 13.6 & 13.6 & - \\
      \hline
       Pass 3 & 7.1 & 2.7 & 1.1 & 3.3 \\
      \hline
       \multicolumn{5}{|c|}{Average number of errors corrected (Z)} \\
      \hline
       & Totals & Sequence 1 & Sequence 2 & Sequence 3 \\
      \hline
       Pass 1 & 5.6 & 5.6 & - & - \\
      \hline
       Pass 2 & 4.4 & 2.2 & 2.2 & - \\
      \hline
       Pass 3 & 1.2 & 0.5 & 0.2 & 0.5 \\
      \hline
  \end{tabular}
  \caption{Cascade stats - average number of errors corrected at each step in cascade. \label{tab.ErrorSequence}}
\end{table}

\begin{table}[htbp]
  \centering
  \begin{tabular}{|c|c|c|}
      \hline
       \multicolumn{3}{|c|}{Average number of errors corrected (X)} \\
      \hline
       BINARY & BICONF & Total \\
      \hline
       65.8 & 1.2 & 67.0 \\
      \hline
       \multicolumn{3}{|c|}{Average number of bits revealed (X)} \\
      \hline
       BINARY & BICONF & Total \\
      \hline
       437.5 & 53.3 & 490.8 \\
      \hline
       \multicolumn{3}{|c|}{Average number of errors corrected (Z)} \\
      \hline
       BINARY & BICONF & Total \\
      \hline
       11.2 & 1.7 & 12.9 \\
      \hline
       \multicolumn{3}{|c|}{Average number of bits revealed (Z)} \\
      \hline
       BINARY & BICONF & Total \\
      \hline
       98.2 & 57.6 & 155.8 \\
       \hline
       \multicolumn{3}{|c|}{Average key block length} \\
      \hline
       X & Z & \\
      \hline
       1,207.7 & 927.2 & \\
       \hline
       \multicolumn{3}{|c|}{Average QBER} \\
      \hline
       X & Z & \\
      \hline
       5.4 & 1.2 & \\
      \hline
       Error correction efficiency (X) & Error correction efficiency (X) & \\
      \hline
       1.31 & 1.59 & \\
      \hline
  \end{tabular}
  \caption{Cascade stats - average number of errors corrected, bits revealed, error correction efficiencies, key block lengths, and QBERs. \label{tab.CascadeStats}}
\end{table}

As was discussed above, the new algorithm allows us to set the desired residual error rate via the parameter $s$ which determines the rate according to $P_{residual} < 2^{-s}$. However, as the residual error rate requirements grow more stringent the efficiency of the error correction algorithm relative to the Shannon limit, $f(x)$, will begin to deteriorate. For all experiments there were no residual errors left in the error corrected key after error correction was performed.

\section{Conclusions}\label{sec.Conclusions}

In conclusion, we have implemented the first experiment to utilize a biased (non-uniform) basis choice in order to increase the efficiency of the number of secure key bits generated from raw key bits. We investigated many of the issues associated with its implementation including choosing the optimal bias, doing a more refined error analysis, and taking care of finite size key effects. We simulated a biased non-polarizing beamsplitter in order to study different biases for the system and their resulting efficiencies. All other aspects of the experiment were implemented in their entirety so that our 50/50 beamsplitter can be easily exchanged with a new fixed non-polarizing beamsplitter with \emph{no} other changes needed to the system. For the near optimal biases of 0.8804/0.1196 (Z/X) for Alice and 0.9062/0.0938 (Z/X) for Bob, we were able to generate 0.4567 secure key bits per raw key bit; whereas, the unbiased case generated 0.2550 secure key bits per raw key bit. This represents an increase in the efficiency of the key generation rate by 79\% over the unbiased case. An improved error correction algorithm was also implemented and statistics for its operation on actual generated key material was discussed.


\section*{Acknowledgements}
Support for this work by NSERC, QuantumWorks, CIFAR, CFI, CIPI, ORF, ORDCF, ERA, and the Bell family fund is gratefully acknowledged. The authors would like to thank N. L\"utkenhaus, R. Kaltenbaek, T. Moroder, H. Hasseler, and O. Moussa for their helpful discussions. We would also like to thank R. Horn and M. Wesolowski for their help with the setup of the experiment. Lastly, the authors would like to thank the anonymous referees for their many comments, which were very useful in improving the quality of this paper


\section*{References}
\bibliographystyle{unsrt}
\bibliography{Paper4_Bibliography}

\end{document}